\documentclass{iopart}
\usepackage{cite}

\def\rmd{{\rm d}} 
\def\beq{\begin{equation}}
\def\eeq{\end{equation}}

\catcode`@=11  
\def\meqalign#1{\null\,\vcenter{\openup\jot\m@th
\ialign{\strut\hfil$\displaystyle{##}$&&$\displaystyle{{}##}$\hfil
\crcr#1\crcr}}\,}
\catcode`@=12  

\def\pmb#1{\setbox0=\hbox{$#1$}%
  \kern-.025em\copy0\kern-\wd0
  \kern.05em\copy0\kern-\wd0
  \kern-.025em\raise.0433em\box0}

\def\bfGamma{\pmb{\Gamma}}
\def\bfmu{\pmb{\mu}} 
\def\bfomega{\pmb{\omega}}  
\def\bfrho{\pmb{\rho}}
\def\bfsigma{\pmb{\sigma}}  
\def\bfOmega{\pmb{\Omega}}  

\jl{6}        
\eqnobysec    

\def\beq{\begin{equation}}
\def\eeq{\end{equation}}

\begin{document}

\title[Spin, Acceleration and Gravity]
{Spin, Acceleration and Gravity}

\author{
Donato Bini${}^{\dag\,\ddag, \P}$,
Christian Cherubini${}^{\ast,||,\ddag}$,
Bahram Mashhoon${}^{\S }$,
}

\address{
  ${}^{\dag}$\
Istituto per le Applicazioni del Calcolo \lq\lq M. Picone\rq\rq, C.N.R.,
   I-- 00161 Roma, Italy
}

\address{
  ${}^{\ddag}$\
  International Center for Relativistic Astrophysics,
  University of Rome, I--00185 Roma, Italy
}

\address{${}^\P$
Sezione INFN di Firenze, Polo Scientifico, Via Sansone 1,
I--50019 Sesto Fiorentino (FI), Italy
}

\address{
  ${}^{\ast}$\
 Faculty of Engineering, University Campus Bio-Medico of Rome, 
via E. Longoni 47, 00155 Rome, Italy.
}

\address{
${}^{||}$\
Institute of Cosmology and Gravitation, University of Portsmouth,
Portsmouth, PO1 2EG, UK}
\address{
  ${}^{\S}$\
  Department of Physics and Astronomy,
University of Missouri-Columbia, Columbia,
Missouri 65211, USA
}

\begin{abstract}
The massless field perturbations of the accelerating Minkowski and Schwarzschild spacetimes are studied. 
The results are extended to the
 propagation of the Proca field in Rindler spacetime. We
examine critically the possibility of existence of a general spin-acceleration coupling in complete analogy with the 
well-known spin-rotation coupling. We argue that such a {\it direct} coupling between spin and linear acceleration does 
not exist. 
\end{abstract}

\section{Introduction}

It was first suggested about forty years ago that polarized objects 
may violate the principle of equivalence of inertial and gravitational 
masses through a coupling of intrinsic spin with the acceleration of 
gravity  proportional to $ \hbar \bfsigma \cdot \mathbf{g} / c$, where 
$\mathbf{S} = \hbar \bfsigma $ is 
the spin of a Dirac particle \cite{M1} and $\mathbf{g}$ is the acceleration of gravity. 
Soon upper limits were placed on 
the strength of such an interaction from the measurements of the 
hyperfine splitting of the ground state of hydrogen \cite{M2} and the 
gravitational acceleration of free neutrons \cite{M3}. 
More generally, an interaction Hamiltonian of the form
\beq
H_{\rm int} = f(r) \hbar \bfsigma\cdot \mathbf{g} / c    
\eeq
has been considered by a number of authors in 
connection with possible violations of parity 
and time-reversal invariance in the gravitational 
interaction \cite{M2, M4,M5,M6}. Over the years, 
experimental claims for the existence of 
such a spin-acceleration coupling have been 
refuted \cite{WR} and at present there is no observational 
evidence in favor of such 
an interaction. Moreover, 
such terms are absent in the usual treatments of the Dirac equation in accelerated systems 
and gravitational fields (see, for instance, \cite{HN,VR1,VR2}). This subject
has been briefly reviewed in \cite{s-a6}. Let us note for the 
sake of concreteness that for a spin-1/2 particle, 
the interaction Hamiltonian $\hbar \bfsigma \cdot \mathbf{g} / c $
implies that the energy difference between the spin-up 
and spin-down states in the Earth's gravitational field 
is $\hbar g_\oplus / c \simeq 2 \times 10^{-23}\,$ eV. 
It is important to recognize that the present 
experimental accuracy for such a measurement 
is an order of magnitude better, i.e. $2 \times 10^{- 24}\,$ eV  \cite{s-a8}. 

In contrast to the coupling of spin with linear acceleration, 
there is ample evidence, both direct and indirect, 
for a general spin-rotation coupling \cite{ s-a6,MNHS,LR}. 
In the case of photons, for instance, helicity-rotation 
coupling has been verified to high accuracy via rotating 
GPS receivers; indeed, this coupling is responsible for 
the phenomenon of phase wrap-up \cite{ASH}. Intrinsic spin 
couples to the rotation of the observer according to the 
Hamiltonian $-\mathbf{S}\cdot\bfOmega$, where $\bfOmega$ is the observer's proper frequency of rotation. 
This phenomenon is ultimately due to the inertia of intrinsic spin. 
Moreover, as a consequence of Einstein's principle of equivalence, 
or equivalently the gravitational Larmor theorem, there is a 
corresponding coupling of intrinsic spin with the gravitomagnetic 
field generated by a rotating gravitational source. This is a direct 
analogue of $-\bfmu \cdot {\mathbf B}$ coupling in electrodynamics extended to the 
gravitational interaction. It follows that in a laboratory fixed 
on the Earth, to every spin Hamiltonian one must add
\beq
\label{eq:2}
\delta H \simeq - \mathbf{S}\cdot \bfOmega_\oplus + \mathbf{S}\cdot \bfOmega_P ,
\eeq
where $c\, \bfOmega_P$ is the gravitomagnetic field of the Earth,
\beq
\bfOmega_P ~= \frac{GJ}{c^2 r^3 }\, [ 3 ( \mathbf{J}\cdot \mathbf{r} ) \mathbf{r} - \mathbf{J} ].    
\eeq
Here $\mathbf{J}$ is the Earth's angular momentum and
 $\bfOmega_P$ is the precession frequency of an ideal test gyroscope held at 
rest in the Earth's gravitational field and $\bfOmega_\oplus$ is the Earth's 
proper rotation frequency.

For a spin-1/2 particle, the spin-rotation part 
of equation (\ref{eq:2}) implies that the maximum energy difference between spin-up and spin-down 
states is $\hbar \Omega_\oplus \simeq 10^{-19}\,$ eV. The experimental results of  \cite{V} can be interpreted to be
an indirect measurement of the spin-rotation coupling for a spin-1/2 
particle \cite{MA}; 
further possible indirect measurements have been discussed by Papini et al. \cite{P1,P2,P3,P4}. Moreover,
 the corresponding energy difference for the spin-gravity term 
in equation (\ref{eq:2}) is $\hbar \Omega_P \simeq 10^{ -29}\,$ eV; the prospects for 
the measurement of this effect have been discussed in \cite{s-a6}. 

  Consider an accelerated observer in Minkowski spacetime. 
The observer carries an orthonormal tetrad frame $\lambda^\mu_{(\alpha)} (\tau) $
along its worldline $x^\mu ( \tau)$, where $\tau $ is the proper 
time along the 
path. For the sake of simplicity, we choose units such 
that $c = G = 1$ in the rest of this paper. The
variation of  the tetrad along the worldline is given by
\beq
    \rmd \lambda^\mu_{(\alpha)} / \rmd \tau = \Phi_{(\alpha)}{}^{(\beta)}  \lambda^\mu_{(\beta)},     
\eeq
where $\Phi_{(\alpha)(\beta)}$ is the antisymmetric acceleration tensor. In analogy 
with the Faraday tensor, one can decompose it into its  \lq\lq electric\rq\rq and  
\lq\lq  magnetic\rq\rq  parts, $\Phi_{(\alpha)(\beta)} \to ( - \mathbf{a}, \bfomega )$, where 
$\Phi_{(0)(i)} = a_{(i)}$ and $\Phi_{(i)(j)} = 1/2\, \epsilon_{ (i)(j)(k)} \omega^{(k)}$.
 The translational acceleration of the worldline is given by
$ A^\mu = \rmd^2 x^\mu / \rmd \tau^2 $ and the projection of this vector on the spatial 
triad results in $a_{(i)} = A_\mu \lambda^\mu_{(i)}$. On the other hand, it can be 
shown that $\bfomega$ is the frequency of rotation of the spatial triad with 
respect to a nonrotating (i.e. Fermi-Walker transported) spatial triad 
along the worldline. Intrinsic spin couples with the  \lq\lq magnetic\rq\rq  part of 
the acceleration tensor and so it is natural to inquire whether a similar 
coupling exists with the  \lq\lq  electric\rq\rq  part of the acceleration tensor. 
Let us note that under the parity transformation, $\mathbf{a}$ behaves as a vector, 
but $\bfomega$ behaves as a pseudovector, while under the time-reversal operation, 
$\mathbf{a}$ remains invariant, but $\bfomega$ changes sign, just as would be expected from 
the electromagnetic analogy. It follows from this discussion that the spin-rotation 
coupling does not violate parity and time-reversal invariance in contrast to the 
coupling of intrinsic spin with translational acceleration.

   In a recent interesting paper, Obukhov \cite{s-a71} has argued that in the treatment 
of Dirac equation in accelerated systems and gravitational fields, the Foldy-Wouthuysen 
transformation is not unique. Taking advantage of this fact, he then showed that 
one can introduce a  $-\mathbf{S}\cdot \mathbf{A} / 2$ term in the low-energy Hamiltonian of a spin-1/2 
particle in a system with uniform acceleration $\mathbf{A}$. It follows from Einstein's 
principle of equivalence that the analogue of this term in the Schwarzschild 
field would be $\mathbf{S}\cdot \mathbf{g} / 2$ \cite{s-a71}. Unless other extra criteria are introduced that 
would rule out such terms, one is left with the possibility that 
terms involving spin-linear acceleration-gravity coupling could in principle 
exist in the treatment of the Dirac equation in linearly accelerated systems 
and gravitational fields. On the other hand, if such an interaction exists for 
a Dirac particle, we expect that it should be a general coupling in analogy 
with the spin-rotation coupling. 
In particular, we note that the spin-linear acceleration-gravity coupling tentatively suggested by Obukhov \cite{s-a71}
does not depend upon the mass of the Dirac particle (assumed to be nonzero in \cite{s-a71}).
Therefore, in section 2 we study the 
propagation of massless fields in a uniformly accelerated system of 
reference and show that a natural coupling of spin with linear 
acceleration does not arise in this case. In section 3, 
we concentrate on the propagation of electromagnetic waves  
and show explicitly that there is no {\it direct} coupling of massless or massive photon 
spin with linear acceleration. 
We reach a similar conclusion in section 4, where we discuss massless field 
perturbations of the vacuum C-metric, which represents the exterior of a uniformly accelerating Schwarzschild spacetime. 
Section 5 contains a discussion of our results.

\section{Massless perturbations of Rindler spacetime}

Imagine a test particle of mass $m$ 
in a background Minkowski spacetime with inertial coordinates $x^\mu=\{t,x,y,z \}$. 
The particle is accelerated along the negative $z-$axis with uniform acceleration $A\ge 0$. 
Let $\tau$ be the proper time of the test particle; then, its worldline can be expressed as
\begin{equation}
\fl\qquad
t= \frac1A \sinh A\tau , \quad x=0, \quad y=0, \quad z=z_0- \frac1A (-1+\cosh A\tau).
\end{equation}
We choose $z_0=-1/A$ for the sake of simplicity; therefore, the path of $m$ is given by
\begin{equation}
x^\mu_m= \frac1A (\sinh A\tau , 0,0, -\cosh A\tau ).
\end{equation}
It is useful to establish a nonrotating (i. e. Fermi-Walker transported) 
tetrad frame along the worldline given by
\begin{eqnarray}
\lambda_{(0)}&=& \cosh A\tau \partial_t - \sinh A\tau \partial_z, \nonumber \\
\lambda_{(1)}&=& \partial_x , \nonumber \\
\lambda_{(2)}&=& \partial_y , \nonumber \\
\lambda_{(3)}&=& - \sinh A\tau \partial_t +\cosh A\tau \partial_z .
\end{eqnarray}
Using this tetrad frame, it is possible to set up a Fermi 
normal coordinate system $\{T,X,Y,Z\}$ along the worldline such that
\begin{equation}
x^\mu -x^\mu_m=X^i \lambda_{(i)}^\mu, \qquad \tau=T.
\end{equation}
It follows that the map $\{t,x,y,z\}\rightarrow \{T,X,Y,Z\}$ can be written as
\begin{eqnarray}\label{COORDYS}
t&=& (\frac1A -Z)\sinh AT, \nonumber \\
x&=&X , \nonumber \\
y&=&Y , \nonumber \\
z&=& -  (\frac1A -Z)\cosh AT.
\end{eqnarray}
Under this coordinate transformation, the line element $\rmd s^2=\rmd t^2-\rmd x^2-\rmd y^2-\rmd z^2$ takes the form
\begin{equation}
\label{I}
\rmd s^2=(1-AZ)^2\rmd T^2-\rmd X^2-\rmd Y^2-\rmd Z^2.
\end{equation}
The coordinates in (\ref{I}) are admissible for $T\in (-\infty , +\infty)$, $X\in (-\infty , +\infty)$, $Y\in (-\infty , +\infty)$ and $Z\in (-\infty , 1/A)$.
In this Rindler spacetime \cite{ri}, the hypersurface-orthogonal Killing vector $\partial_T$ is timelike, but becomes null on the horizon $Z=1/A$.
The coordinate system breaks down beyond this limit and is not admissible; in fact $Z=1/A$ corresponds to a null cone in  $\{t,x,y,z\}$ coordinates, 
since  $z^2-t^2=(Z-1/A)^2$.

Let us consider the metric (\ref{I}) with the associated Newman-Penrose (NP) principal null frame 
\begin{eqnarray} 
\label{NPtetrad}
\mathbf{l}&=&\frac{(1-AZ)^{-2}}{\sqrt{2}}[\partial_T-(1-AZ)\partial_Z],\nonumber \\
\mathbf{n}&=&\frac{1}{\sqrt{2}}[\partial_T+(1-AZ)\partial_Z],\nonumber \\
\mathbf{m}&=&\frac{1}{\sqrt{2}}[\partial_X+i\partial_Y];
\end{eqnarray}
the only nonvanishing NP spin coefficient is $\gamma=A/\sqrt{2}$, while the Weyl scalars are obviously all zero.
We do not wish to deviate from the standard NP notation; therefore,
it is important to observe that the notation employed in this section, section 4 and appendix A is partly independent of the rest of this paper.
Using the standard NP terminology and notation \cite{chandra}, the massless perturbations of the flat spacetime (\ref{I}) are described by:
\begin{eqnarray}
\fl
\quad\label{mia1}
&&
\{[D-\rho^{*}+\epsilon^*+\epsilon-2s(\rho +\epsilon)](\Delta+\mu-2 s \gamma)
\\ \fl\quad
&&
  -[\delta+\pi^{*}-\alpha^{*}+\beta
-2 s(\tau+\beta)] \,(\delta^{*}+\pi-2 s\alpha)
  -2(s-1)(s-1/2)\psi_{2}\}\Psi_s=0 
\nonumber
\end{eqnarray}
for spin weights $s=1/2,1,2$  and
\begin{eqnarray}\label{mia2}
\fl\quad
&&
\{[\Delta-\gamma^{*}+\mu^{*}-\gamma-2 s (\gamma+\mu)](D-\rho-2s\epsilon)
\\ \fl\quad
&&
-[\delta^{*}-\tau^{*}+\beta^{*}-\alpha
-2 s (\alpha+\pi)](\delta-\tau-2 s \beta)
-2(s+1)(s+1/2)\psi_{2}\}\Psi_s=0
\nonumber
\end{eqnarray}
for $s=-1/2,-1,-2$. The case $s=\pm 3/2$ can be derived instead  by following the work of 
G\"uven \cite{Guven}, which is expressed in the alternative Geroch-Held-Penrose formalism \cite{GHP}. Finally the case $s=0$ is given by
\begin{eqnarray}
\label{mia3}
\fl\quad
\meqalign{
&[D\Delta+\Delta D-\delta^* \delta-\delta\delta^*
+(-\gamma-\gamma^*+\mu+\mu^*)D+(\epsilon+\epsilon^*-\rho^*-\rho)\Delta \cr
&+(-\beta^*-\pi+\alpha+\tau^*)\delta+(-\pi^*+\tau-\beta+\alpha^*)\delta^*]\Psi_s=0\ .
\cr}
\end{eqnarray}

\begin{table} [h]
\begin{tabular}{c||ccccccccc}
$s\quad$ & $0\quad$ & $1/2\quad$ & $-1/2\quad$ & $1\quad$ & $-1\quad$ & $3/2\quad $& $-3/2\quad$ & $2\quad $ & $ -2\quad $\\
$\Psi_s\quad$ & $\Phi\quad$ & $\chi_0\quad$ & $\chi_1\quad$ & $\phi_0\quad $ & $\phi_2\quad$ &
$\Omega_0\quad$ & $\Omega_3\quad$ & $\psi_0\quad$ & $\psi_4\quad$ 
\end{tabular}
\caption{The spin-weight $s$ and the physical field component $\Psi_s$ for the master equation in the Minkowski spacetime.}
\label{TAB}
\end{table}
All these equations can be cast in a unique \lq\lq master equation\rq\rq \cite{teuk1,teuk2}
\begin{eqnarray}\label{TME1} 
&&\left[ 
(1-AZ)^{-2}\frac{\partial^2}{\partial {T^2}}-\frac{\partial^2}{\partial {X^2}}-\frac{\partial^2}{\partial {Y^2}}- 
\frac{\partial^2}{\partial {Z^2}}\right. \nonumber \\
&& \left. -2As(1-AZ)^{-2}\frac{\partial}{\partial {T}}+A(2s+1)(1-AZ)^{-1}\frac{\partial}{\partial {Z}}
\right]\Psi_s=0 . 
\end{eqnarray} 

The quantities satisfying the master equation are listed in table \ref{TAB}. The transformation properties of the master equation (\ref{TME1}) are discussed in appendix A.
Normal modes of equation (\ref{TME1}) can be expressed as $\Psi_s(T,X,Y,Z)=e^{-i\omega T}e^{i k_X X}e^{i k_Y Y}P_s(Z)$, with $P_s(Z)$ satisfying the equation
\begin{equation}
\label{eq:66}
P_s^{''}-\frac{A(1+2s)}{(1-AZ)}P_s^{'}-[k_X^2+k_Y^2-\frac{\omega^2-2isA\omega}{(1-AZ)^2}]P_s=0 ,
\end{equation}
where the prime denotes differentiation with respect to $Z$.
If $A = 0$, then (\ref{eq:66}) has the expected solutions $e^{\pm ik_Z Z}$, where $\omega^2 = k_X^2 + k_Y^2 + k_Z^2$. For $A > 0$, it is useful to assume that $k_X^2+k_Y^2>0$ and consider the following transformations in (\ref{eq:66})
\begin{equation}
\label{eq:67}
Z=\frac1A-\frac{iq}{\sqrt{k_X^2+k_Y^2}},\qquad
P_s(Z)=(iq)^{-s}Q_s(q),
\end{equation}
which result in
\begin{equation}
\label{eq:69}
q^2 \ddot Q_s + q\dot Q_s +(q^2-\sigma^2)Q_s=0.
\end{equation}
Here $\sigma=s+i\frac{\omega}{A}$ and (\ref{eq:69}) is the standard Bessel equation; therefore, the general solution of (\ref{eq:69}) is
\begin{equation}
\label{69}
Q_s(q)=C_1 J_\sigma (q)+C_2 N_\sigma (q), 
\end{equation}
where $J_\sigma$ and $N_\sigma$ are the standard Bessel and Neumann functions, respectively, and $C_1$ and $C_2$ are constants.

The coordinates in (\ref{I}) are admissible for $Z< 1/A$; therefore, it follows from (\ref{eq:67}) that the admissible range for $q$ is such that $iq$ is a positive real number. The constants in (\ref{69}) are determined by the boundary conditions that would be appropriate for the particular physical situation under consideration. Inspection of the Bessel and Neumann functions reveals that there is no {\it direct} coupling between the spin-weight $s$ and the acceleration $A$. For instance,

\begin{eqnarray}
J_\sigma(q)&=& (q/2)^\sigma \sum_{k=0}^\infty \frac{(-q^2/4)^k}{k!\,  \Gamma (\sigma +k+1)}, \nonumber \\
N_\sigma(q)&=&\frac{J_\sigma(q)\cos \sigma \pi-J_{-\sigma} (q)}{\sin \sigma \pi}
\end{eqnarray}
by definition and 
\begin{equation}
\label{eq:75}
(q/2)^\sigma=(q/2)^s \left[\cos (\frac{\omega}{A}\ln \frac{q}{2})+i \sin (\frac{\omega}{A}\ln \frac{q}{2}) \right]. 
\end{equation}
A similar functional separation between the real and imaginary parts of $\sigma$  is also present in the gamma function in $J_\sigma(q)$; hence, the spin part does not directly couple with the $\omega /A$ part. An analogous argument holds for the Neumann function as well.

Finally let us consider the $k_X=k_Y=0$ case that we excluded from the treatment in (\ref{eq:67})-(\ref{eq:75}). Thus for $A>0$ let us define in this case $\hat q$ and $\hat Q_s$ such that

\begin{equation}
Z=\frac1A-\hat q, \qquad P_s(Z)=\hat q{}^{-s}\hat Q_s(\hat q).
\end{equation}
It follows from (\ref{eq:66}) that
\begin{equation}
\hat q{}^2 \frac{\rmd^2 \hat Q_s}{\rmd \hat q{}^2}+\hat q \frac{\rmd \hat Q_s}{\rmd \hat q}-\sigma^2 \hat Q_s=0.
\end{equation}
The general solution of this equation can be expressed as
\begin{equation}
\hat Q_s =\hat C_1 \hat q{}^\sigma + \hat C_2 \hat q{}^{-\sigma} ,
\end{equation}
where $\hat C_1$ and $\hat C_2$ are constants. It is now straightforward to conclude that there is no direct coupling between spin and acceleration in the perturbing massless fields. This issue is further discussed in the next section.

\section{Electromagnetic waves}

The purpose of this section is to show explicitly that there is no coupling between helicity and linear acceleration by considering electromagnetic radiation as a perturbation on Rindler and Schwarzschild spacetimes.

To describe the propagation of electromagnetic waves in a spacetime background with Cartesian coordinates $X^\mu=(T,\mathbf{X})$ and a metric
$\rmd s^2= g_{\mu\nu}\rmd X^\mu \rmd X^\nu$ 
with signature +2, it is useful to employ an approach based on the idea that the gravitational field may be replaced by an effective medium with special constitutive properties \cite{s-a1,s-a2}.
To this end, we first write the source-free Maxwell equations as $F_{[\mu\nu,\rho]}=0$ and $(\sqrt{-g}F^{\mu\nu})_{,\nu}=0$.
Next, we introduce the standard decompositions $F_{\mu\nu} \to (\mathbf{E},\mathbf{B})$ and $\sqrt{-g}F^{\mu\nu}\to (-\mathbf{D},\mathbf{H})$, so that in terms of $\mathbf{E}$, $\mathbf{B}$, $\mathbf{D}$ and  $\mathbf{H}$, Maxwell's equations take the form of the standard electromagnetic field equations in inertial coordinates but in the presence of an effective \lq\lq material\rq\rq  medium
\beq\label{eq:31}
\fl\qquad\,
\nabla \cdot \mathbf{D}=0, \quad \nabla \cdot \mathbf{B}=0,\quad \nabla \times \mathbf{E}=-\partial_T \mathbf{B},\quad
\nabla \times \mathbf{H}=\partial_T \mathbf{D}.
\eeq
In this case, the constitutive relations for the gravitational medium are
\beq\label{eq:32}
D_i=\epsilon_{ij}E_j-(\bfGamma \times \mathbf{H})_i, \quad
B_i=\mu_{ij}H_j+(\bfGamma \times \mathbf{E})_i,
\eeq
where $\epsilon_{ij}=\mu_{ij}=-\sqrt{-g}g^{ij}/g_{00}$ and $\Gamma_i=-g_{0i}/g_{00}$. This approach, 
originally due to Skrotskii \cite{S,P,dF}, is manifestly invariant under conformal transformations of the metric tensor.

The linearity of the field equations (\ref{eq:31}) and (\ref{eq:32}) implies that we can treat the fields as complex, such that
their real parts would correspond to the actual physical fields. 
Using complex fields, we now define the Riemann-Silberstein \cite{HO} fields $\mathbf{F}^\pm= \mathbf{E}\pm i \mathbf{H}$ and $\mathbf{S}^\pm= \mathbf{D}\pm i \mathbf{B}$;
Maxwell's equations then take the form
\beq\label{eq:33}
\fl\qquad
\nabla \cdot \mathbf{ S}^\pm=0, \quad \nabla \times \mathbf{F}^\pm=\pm i \partial_T \mathbf{S}^\pm, \quad
\mathbf{S}^\pm=\epsilon \mathbf{F}^\pm \pm i \bfGamma \times \mathbf{F}^\pm .
\eeq
The field amplitudes $\mathbf{F}^+$ and $\mathbf{S}^+$ (with $\mathbf{F}^-=\mathbf{S}^-=0$) describe the propagation of positive-helicity
waves, while $\mathbf{F}^-$ and $\mathbf{S}^-$ (with $\mathbf{F}^+=\mathbf{S}^+=0$) describe the propagation of negative-helicity waves. The concept of helicity 
in accelerated systems and gravitational fields is simply an extension of this standard 
notion for fields in Minkowski spacetime. Equations (\ref{eq:33})  completely decouple, so that $\mathbf{F}^+$ and $\mathbf{S}^+$ scatter independently of 
$\mathbf{F}^-$ and $\mathbf{S}^-$ in this linear perturbation analysis.

Consider the propagation of electromagnetic waves in the Rindler spacetime. We find that $\bfGamma=0$ and
\beq
\label{eq:34}
\epsilon_{ij}=\mu_{ij}=N_A\delta_{ij}, \qquad N_A=1/(1-AZ),
\eeq
where $N_A$ has the interpretation of index of refraction. It diverges on the horizon and for $Z\in (-\infty, 1/A)$, $N_A$ ranges from $0$ to infinity. The Rindler spacetime is static; therefore, we can assume a time dependence of the form $e^{-i\omega T}$ in (\ref{eq:33}). The field equations then reduce to
\beq\label{eq:35}
\nabla \times \mathbf{F}^\pm=\pm \omega N_A \mathbf{F}^\pm .
\eeq
To see if a helicity coupling of the form $\mathbf{S} \cdot \mathbf{A}$ exists, it is natural to focus attention on the propagation of circularly polarized waves along the $Z-$axis. We find that plane-wave solutions of (\ref{eq:35}) exist for circularly polarized waves of the form
\beq
\label{eq:36}
\mathbf{F}^\pm=f_0^\pm (\hat\mathbf{X}\pm i\hat\mathbf{Y})e^{-i\omega (T-Z^*)},
\eeq
where $f_0^\pm$ are constant amplitudes and $Z^*$ is the tortoise coordinate for Rindler spacetime given by
\beq\label{eq:37}
Z^*=\int_0^Z \frac{\rmd Z'}{1-AZ'}=-\frac{1}{A}\ln (1-AZ).
\eeq
Thus $Z^*\in (-\infty, \infty)$ for $Z\in (-\infty, \frac1A)$, $Z^*=0$ when $Z=0$ and $Z^*\to Z$ as $A\to 0$.
The $\mathbf{S} \cdot \mathbf{A}$ coupling is manifestly absent in equation (\ref{eq:36}), so that the plane of polarization
of an initially linearly polarized beam does not rotate as the wave propagates along the direction of acceleration. The phenomenon 
of optical activity is thus absent in this "medium", as is already clear from the fact that 
the index of refraction is independent of the state of circular polarization of the wave. The
result is consistent with previous studies of electrodynamics in a linearly accelerated frame of reference via a different approach \cite{s-a3}.
Moreover the influence of linear acceleration on the optical response of matter has been studied in \cite{s-a4,s-a5} and the main effect of light propagation through linearly accelerating media is that photons propagating parallel to the direction of acceleration have higher energies upon transit.

It is interesting to extend the above analysis to the 
case of massive photons described by the Proca equations \cite{PR}. 
Obukhov's treatment \cite{s-a71} involved a massive Dirac particle; 
therefore, it is important to demonstrate that our results 
for the massless photon extend to the massive photon as well. The Proca equations are
\beq
\nabla^\alpha F_{\beta\alpha}+\mu^2 \mathcal{A}_\beta =4\pi J_\beta ,\qquad F_{\alpha\beta}=\partial_\alpha \mathcal{A}_\beta - \partial_\beta \mathcal{A}_\alpha ,
\eeq
where $\hbar\mu$ is now the mass of the photon.
Introducing $\mathbf{F}^{\pm}$ and $\mathbf{S}^\pm$ as above and defining
\beq
\sqrt{-g}J^\mu \,\to\, (\hat \rho, \mathbf{j})
\eeq
and
\beq
\sqrt{-g}\mathcal{A}^\mu \,\to\, (\Phi, \pmb{ \mathcal{A}}),
\eeq
the Proca equations can be cast in the form
\begin{eqnarray}
&& \nabla \times \mathbf{F}^{\pm}=\pm i \partial_T \mathbf{S}^{\pm} \pm 4\pi i [\, \mathbf{j}-\frac{\mu^2}{4\pi}\pmb{ \mathcal{A}}\, ], \nonumber \\
&& \nabla \cdot \mathbf{S}^\pm =4\pi (\hat \rho-\frac{\mu^2}{4\pi} \Phi),
\end{eqnarray}
which, together with the constitutive relations, 
are consistent with charge conservation once the 
Lorentz condition ($\nabla \cdot \pmb{ \mathcal{A}} + \partial_T \Phi = 0$) is 
imposed on the vector potential. In the absence of 
electromagnetic sources, the case in which we are 
interested here, and in terms of complex fields, the Proca equations become
\begin{eqnarray}
\label{eqs1}
&& \nabla \times \mathbf{F}^{\pm}=\pm i \partial_T \mathbf{S}^{\pm} \mp i\mu^2 \pmb{ \mathcal{A}}, \nonumber \\
&& \nabla \cdot \mathbf{S}^\pm =- \mu^2 \Phi ,
\end{eqnarray}
to which one must add here  the Lorentz condition $\nabla_\alpha \mathcal{A}^\alpha=0$, i. e. $\nabla \cdot \pmb{ \mathcal{A}} +\partial_T \Phi=0$, as compatibility condition now 
and not a gauge choice. 
The constitutive relations connecting $\mathbf{S}^\pm$ to $\mathbf{F}^\pm$ remain the same as in the massless 
case above (cf. equation (\ref{eq:33}).

Let us now specialize these equations to the static Rindler spacetime and look for a transverse plane wave solution propagating along the $Z-$direction. We expect that for $\mu=0$ the solution would reduce to eq. (\ref{eq:36}). Specifically we assume that
\begin{eqnarray}
\mathbf{F}^{\pm} &=& f_0^\pm W(Z) (\hat \mathbf{X}\pm i \hat \mathbf{Y}) e^{-i\omega T}, \nonumber \\
\pmb{ \mathcal{A}}^{\pm} &=& f_0^\pm U(Z) (\hat \mathbf{X}\pm i \hat \mathbf{Y}) e^{-i\omega T}, 
\end{eqnarray}
where $f_0^\pm$ are constants as before. It follows from the compatibility condition that $\Phi=0$.
Eq. (\ref{eqs1}) implies in this case that
\beq
\label{eq:313}
\frac{\rmd W}{\rmd Z}= \frac{i\omega W}{1-AZ}+\mu^2 U,
\eeq
while the relation connecting $F_{\mu\nu}$ to the vector potential $\mathcal{A}_\mu$ implies that
\beq
\label{eq:314}
W=\frac{i\omega U}{1-AZ}+(1-AZ)\frac{\rmd }{\rmd Z} \left( \frac{U}{1-AZ} \right) .
\eeq
In deriving eq. (\ref{eq:314}) we have made use of the fact that $\sqrt{-g}=1-AZ=N_A^{-1}$, $\mathbf{D}=N_A \mathbf{E}$, 
$\mathbf{B}=N_A \mathbf{H}$ and $\mathbf{S}^\pm =N_A \mathbf{F}^\pm$.
Substituting $W$ from eq. (\ref{eq:314}) into eq. (\ref{eq:313}), we find
\beq
\label{eq:315}
\frac{\rmd^2 U}{\rmd Z^2}+\frac{A}{1-AZ}\frac{\rmd U}{\rmd Z}+\left[\frac{A^2+\omega^2}{(1-AZ)^2}-\mu^2 \right] U=0.
\eeq
If $A = 0$, 
this equation has solutions of the form ${\rm exp}(\pm ikZ)$, where $k=\sqrt{\omega^2 - \mu^2}$. 
These correspond to the standard matter waves propagating along the $Z-$direction. 
Therefore, in the rest of this discussion we assume that $A > 0$. 
Let us note that for $\mu = 0$, equation (\ref{eq:315}) has
the solution $U=(2i\omega)^{-1}(1-AZ){\rm exp}(i\omega Z^\ast)$ and the corresponding $W$ is $W(Z)={\rm exp}(i\omega Z^\ast)$ as expected. We therefore look for a solution of equation 
(\ref{eq:315}) of the form
\beq
U(Z)=\frac{(1-AZ)}{2i\omega}e^{i\omega Z^\ast}\chi(Z).
\eeq
The result is 
\beq
\label{eq:317}
\chi(Z)=e^{-\frac{\mu (1-AZ)}{A}}\, F\left(\frac12 -i\frac{\omega}{A}, 1-2i\frac{\omega}{A}, 2\mu \frac{1-AZ}{A}\right),
\eeq
where $F$ is the confluent hypergeometric function
\beq
F=1+\left(\mu \frac{1-AZ}{A}\right)+\frac34 \frac{A-\frac23i\omega}{A-i\omega}\left(\mu \frac{1-AZ}{A}\right)^2+\ldots
\eeq
It is straightforward to relate this function to the modified Bessel function of the first kind.
Moreover,  solution (\ref{eq:317}) depends on the frequency of the wave but is independent of its state of polarization. Therefore, as before, the plane of polarization of an initially linearly polarized Proca wave does not rotate as the plane wave propagates along the acceleration direction. 
In addition to these transverse waves, 
the Proca equations (\ref{eqs1}) have solutions 
corresponding to the propagation of longitudinal 
waves along the $Z-$direction. We do not consider 
them here as they are not directly relevant to the spin-acceleration coupling.

Let us now consider the propagation of electromagnetic waves in the Schwarzschild spacetime. 
In the rest of this paper, we assume that the photon is massless; 
the treatment of Proca's equations in gravitational fields is beyond the scope of this work. The Maxwell equations in isotropic coordinates have the form (\ref{eq:35}), except that $N_A\to N_S$, where
\beq\label{eq:38}
N_S=\left(1+\frac{2m}{\rho}\right)^3\left(1-\frac{2m}{\rho}\right)^{-1}
\eeq
is the index of refraction for the exterior Schwarzschild field, $\rho$ is the isotropic radial coordinate and $m$ is the mass of the black hole.
As before, $N_S$ diverges at the horizon $\rho=m/2$; moreover, it decreases monotonically toward unity as $\rho$ increases toward infinity.
The scattering of electromagnetic radiation from a black hole has been treated at length in \cite{s-a1,s-a2}; in particular, it has been shown that the polarization of the incident electromagnetic radiation is unaffected by a Schwarzschild source due to its spherical symmetry.
It follows that the $\mathbf{S} \cdot \mathbf{g}$  coupling is also nonexistent for photons.

Let us finally consider the propagation of electromagnetic waves in the accelerating Schwarzschild spacetime given by the vacuum C-metric. This metric is described in appendix B.
It can be shown \cite{kin69,kinwal} that the C-metric is a nonlinear superposition of the Schwarzschild and Rindler metrics. The C-metric, in the regime under consideration $(3\sqrt{3}mA<1)$ is static. Though the presence of the uniform acceleration $A$ breaks the spherical symmetry of the Schwarzschild spacetime, it does so in the effective index of refraction of the gravitational medium. To see this clearly, we consider the linearized C-metric in isotropic coordinates
\beq
\label{eq:39}
\rmd s^2=(1-\frac{2m}{\rho}-2AZ)\rmd T^2-(1+\frac{2m}{\rho})(\rmd X^2+\rmd Y^2+\rmd Z^2),
\eeq 
where terms of order $m^2,mA,A^2$ and higher have been neglected. The wave propagation equations are again of the form (\ref{eq:35}) with $N_A \to N_C$,
\beq\label{eq:310}
N_C \sim 1+\frac{2m}{\rho}+AZ
\eeq
in the linear approximation under consideration here. This again indicates that there is no direct helicity-acceleration coupling for photons.
Massless field perturbations of the C-metric
can be studied along the lines of section 2. In coordinates that easily lend themselves to physical interpretation, such as the one we have used thus far, it is not possible to separate the field equations for the C-metric (or its linear version).
On the other hand, it is possible to achieve such a separation in certain physically obscure coordinate systems \cite{prest}.
This is presented in the following section using an approach that is simpler than the one followed in \cite{prest}.

\section{Massless perturbations of accelerating Schwarzschild spacetime}

A master equation, analogous to the one derived in the Kerr spacetime \cite{Guven,teuk1, teuk2, Detweiler, Wainwright, RARITA2,Finley2} and describing massless field perturbations of any spin, has been recently studied in \cite{prest} on the C-metric background. 
However the physical content of this equation  is not yet completely understood,  because  the master equation cannot be integrated exactly but only separated in $\{t,\tilde x,\tilde y,\tilde z \}$ coordinates that are described in appendix B.

We present the master equation  for the C-metric in a slightly different form compared with the one obtained by Prestidge \cite{prest}.
In fact, we use here a principal NP frame which is also Kinnersley-like, i. e. it has the  NP spin coefficient $\epsilon=0$. 
This allows some further simplification and puts this development in a form very close to the black hole case, where  the master equation formalism has been successfully developed.

With the C-metric in the form (\ref{met_txyz}), a Kinnersley-like NP principal null tetrad can be easily constructed
with 
\begin{eqnarray}
\mathbf{l}&=& A(\tilde x +\tilde y)^2  \left(\frac{1}{\tilde F} \partial_t + \partial_{\tilde y}\right),\qquad \nonumber \\
\mathbf{n}&=&  \frac{A}{2} \left( \partial_t -\tilde F\partial_{\tilde y}\right ) ,\qquad
 \nonumber \\
\mathbf{m}&=&\frac{\tilde G^{1/2}A(\tilde x+ \tilde y)}{\sqrt{2}} \left(\partial_{\tilde x} +\frac{i}{\tilde G}\partial_{\tilde z}\right) .
\end{eqnarray}

The nonvanishing spin coefficients are
\begin{eqnarray}
\fl\quad \mu &=&\frac{A^2 \tilde F}{2\rho},\quad  
\tau =\frac{A}{\sqrt{2}}\tilde G^{1/2}=-\pi, 
\quad \rho=A(\tilde x + \tilde y), \quad \beta =\frac{\rho}{4\sqrt{2}} \frac{\tilde G'}{\tilde G^{1/2}},
\nonumber \\
\fl\quad \alpha &=&\frac{A\tilde G^{-1/2}}{4\sqrt{2}}[\tilde G -\tilde y \tilde G' +3-\tilde x^2],\quad 
\gamma =\frac{A}{4 (\tilde x+ \tilde y)}[\tilde F +\tilde x \dot {\tilde F} +3-\tilde y^2],
\end{eqnarray}
while the only surviving Weyl scalar is $\psi_2=-mA^3(\tilde x+\tilde y)^3$; here a prime and a dot denote differentiation with respect to $\tilde x$  and $\tilde y$, respectively.
Following the approach of Prestidge \cite{prest}, rescaling the unknown $\psi_s$ of the master equation (for the various $\psi_s$ satisfying the master equation, see e.g. \cite{KNTN}) we find that
\begin{equation}
\label{psiteuk}
\psi_s=(\tilde x+\tilde y)^{(2s+1)}e^{-i\omega t}e^{ik_3\tilde z}X_s(\tilde x) Y_s(\tilde y) 
\end{equation}
gives separated equations for $X_s$ and $Y_s$:
\begin{eqnarray}
\label{eqxy}
\fl
&&  X_s''+\frac{\tilde G '}{\tilde G}  X_s'+
[\frac{-4S-s^2+2p  \tilde x (s^2-4)}{4\tilde G} \nonumber \\
\fl&&  \qquad -
\frac{(-24 p  k_3+ s) s\tilde x^2+2s(9\d  s -4k_3)\tilde x  +3s^2+4k_3^2}
{4\tilde G^2}]X_s=0, \nonumber \\
\fl && \ddot Y_s +\frac{\dot {\tilde F} (s+1)}{\tilde F}  \dot Y_s+
[\frac{S+s(s+1)-2p  \tilde y (s+1)(2s+1)}{\tilde F}+
\frac{\omega (\omega-is\dot{ \tilde F})}
{\tilde F^2}] Y_s=0, 
\end{eqnarray}
where  $S$ is a separation constant and $p =mA$ is a dimensionless parameter.
Because of the symmetry of the metric under the exchange of $\tilde x$  and $\tilde y$, one would expect a similar property to hold for these two equations. It can be shown that this is exactly the case (modulo further replacement of $\tilde y\to -\tilde x$, $\omega\to ik_3$, $s\to -s$) when one uses  the following  rescaling for $X_s(\tilde x)$ and $Y_s(\tilde y)$
\begin{equation}
X_s(\tilde x) \to  \frac{X_s(\tilde x)}{\tilde G^{1/2}}, \qquad Y_s(\tilde y) \to  \frac{Y_s(\tilde y)}{\tilde F^{(s+1)/2}}. 
\end{equation}
Thus, without any loss of generality one can consider the equation for $X_s$ only. 
This equation, in turn, cannot be solved exactly, unless $p =0$.
In this limit, with $\tilde x=\cos\theta$, one gets
\begin{equation}
\frac{\rmd^2 X_s}{\rmd \theta^2}+\cot \theta  \frac{\rmd X_s}{\rmd \theta}- \left[S+\frac{s^2-2k_3s\cos \theta +k_3^2}{\sin^2\theta}\right]X_s=0,
\end{equation}
so that with $S=-l(l+1)$ and $\tilde z=\phi$, it is easy to show that $X_s(\tilde x)e^{ik_3\tilde z}$ reduces to the standard spin-weighted spherical harmonics. 

Let us consider then the equation for $X_s$ in (\ref{eqxy}), where we set $\tilde x=\cos\theta$ and use the rescaling 
\begin{equation}
X_s(\theta)=\frac{\sin \theta}{\tilde G{}^{1/2}}\mathcal{T}_s(\theta).
\end{equation}
The equation for $\mathcal{T}_s$ is then
\begin{equation}
\label{eq:A8}
\frac{\rmd^2 \mathcal{T}_s}{\rmd \theta^2}+\cot \theta  \frac{\rmd \mathcal{T}_s}{\rmd \theta}- \mathcal{V}\mathcal{T}_s=0,
\end{equation}
where  $\mathcal{V}$ is given by
\begin{equation}
\mathcal{V}=\frac{1}{(1-2p \cos \theta \cot^2 \theta)^2}\left[
p^2 \mathcal{V}_{(2)}(\theta)  +p  \mathcal{V}_{(1)}(\theta) +\mathcal{V}_{(0)}(\theta)
\right]
\end{equation}
and the coefficients
\begin{eqnarray}
\fl\quad \mathcal{V}_{(2)}(\theta)&=& (1-s^2)\cos^2\theta -(1+s^2)\cot^2 \theta +4\cot^6 \theta,\nonumber \\
\fl\quad \mathcal{V}_{(1)}(\theta) &=&2 \cos \theta \left[ 2s^2 (1+\cot^2\theta)-S\cot^2\theta-2(1+\cot^2\theta)^2\right]-6k_3s\cot^2\theta ,\nonumber \\
\fl\quad \mathcal{V}_{(0)}(\theta)&=&  S+\frac{s^2-2k_3 s \cos \theta +k_3^2}{\sin^2\theta} ,
\end{eqnarray}
do not depend on $p$. We recall that in the case under 
consideration here $p < 1/ (3\sqrt{3})$. For $p \ll 1$, it is straightforward to develop a perturbation  series solution to equation (\ref{eq:A8}) in powers of $p$.
In this way, 
terms of the form $ps = msA$ and higher order appear in $\mathcal{V}$, but a
{\it direct} spin-acceleration coupling term $sA$ 
that would be independent of mass $m$ does not exist in $X_s$ and hence $\psi_s$;
 therefore, we may conclude that this coupling does not exist.

\section{Discussion}

For fermions, the idea of spin - acceleration coupling has a long history that has been reviewed in \cite{s-a6}. 
Previous suggestions for the existence of the coupling of the spin with the Earth's gravitational acceleration have been experimentally ruled out, but such an interaction has been tentatively resurrected by Obukhov \cite{s-a71, s-a72, s-a73} in his
recent discussion of the observable consequences of the Dirac equation in accelerated systems and gravitational fields.
Obukhov \cite{s-a71} has employed exact Foldy - Wouthuysen (\lq\lq FW") transformations to decouple the positive and negative energy states and obtained in this way an interaction term of the form $- \mathbf{S} \cdot \mathbf{A} /2$ in a system with uniform linear acceleration $\mathbf{A}$.
The analogue of this term in the Earth's gravitoelectric field would be $\mathbf{S} \cdot \mathbf{g} /2$, and the energy difference between the states of a Dirac particle with spin polarized up and down in the laboratory would therefore be $\hbar g_\oplus /2 \simeq 10^{-23}$ eV. This energy is larger than what can be detected by present experimental techniques by a factor of five \cite{s-a8}. Thus the presence of this spin-acceleration term in the Hamiltonian could in principle be ruled out by observation. On the other hand, the exact FW transformation has a basic ambiguity, namely, it is defined up to a unitary transformation. Thus a unitary transformation may be employed to eliminate the spin-acceleration term as demonstrated by Obukhov \cite{s-a71}. 
It is therefore not evident from \cite{s-a71} what one can predict to be the observable consequences of the Dirac theory in accelerated systems and gravitational fields. 
This general problem requires further investigation; however, we have limited our considerations here to Obukhov's spin-acceleration terms that are independent of the mass of the Dirac particle \cite{s-a71}. We have therefore investigated the propagation of massless fields with spin in Rindler and Rindler-Schwarzschild  
spacetimes as well as the propagation of the Proca field in Rindler spacetime and
find no theoretical evidence for the spin-acceleration-gravity coupling.
It must be remarked, 
however, that our conclusions are based on minimally 
coupled fields as well as the absence of any locally Lorentz-violating interactions \cite{OBp}.

Our theoretical arguments notwithstanding, it would be very interesting 
to verify experimentally the absence of spin-acceleration coupling for fermions.
An upper limit on 

\begin{equation}
\frac12 \mathbf{S} \cdot \mathbf{g} = \left(\frac{M_\oplus}{2\rho_\oplus^3}\right) \mathbf{S} \cdot \bfrho_\oplus
\end{equation}
would also be useful in limiting the magnitude of the Earth's gravitomagnetic monopole moment $\hat \mu$.
This follows from the fact that the coupling of
spin with the gravitomagnetic monopole moment is given by \cite{bcjm}
\begin{equation}
 \left(\frac{\hat \mu{}_\oplus}{\rho_\oplus^3}\right) \mathbf{S} \cdot \bfrho_\oplus\, ,
\end{equation}
where $-\hat \mu{}_\oplus$ is the NUT parameter of the Earth;
other aspects of the NUT parameter including possible ways of placing limits on it have been discussed in \cite{KNTN, bcjm}
and references cited therein.

\section*{Acknowledgments}
BM is grateful to Friedrich Hehl and Yuri Obukhov for helpful discussions and correspondence.

\appendix
\section {Master equation in Rindler spacetime}

In section 2, the master equation (\ref{TME1}) in flat spacetime has been discussed in the accelerated coordinate system $\{T,X,Y,Z\}$ and the normal modes have been  obtained.
To check the consistency of our formalism, equation  (\ref{TME1}) can be transformed by passing to the standard (non-accelerating) Cartesian Minkowski coordinates. In fact, 
inverting the relations (\ref{COORDYS}) we obtain 
\begin{equation} 
\fl\quad
T=\frac{1}{2A}\ln\left(\frac{z-t}{z+t}\right)\,,\quad X=x\,,\quad Y=y \,,\quad Z=\frac{1}{A}-\sqrt{z^2-t^2}\,. 
\end{equation} 
The metric (\ref{I}) then is $\rmd s^2=\rmd t^2-\rmd x^2-\rmd y^2-\rmd z^2$ and the corresponding NP tetrad (\ref{NPtetrad}) becomes
\begin{eqnarray} 
\label{np2}
\mathbf{l}&=&\frac{-1}{\sqrt{2}A(z+t)}[\partial_t-\partial_z],\nonumber \\ 
\mathbf{n}&=&\frac{-A(z+t)}{\sqrt{2}}[\partial_t+\partial_z],\nonumber \\ 
\mathbf{m}&=&\frac{1}{\sqrt{2}}[\partial_x+i\partial_y]. 
\end{eqnarray} 
The master equation (\ref{TME1}) then results in
\begin{equation}\label{TME2} 
\left[ 
\frac{\partial^2}{\partial {t^2}}-\frac{\partial^2}{\partial {x^2}}-\frac{\partial^2}{\partial {y^2}}- 
\frac{\partial^2}{\partial {z^2}}+\frac{2s}{(z+t)}\left( \frac{\partial}{\partial {t}} -\frac{\partial}{\partial{z}} \right)
\right]\Psi_s=0\, . 
\end{equation} 
The null NP frame (\ref{np2}) still manifests anholonomy because of the only nonvanishing (new) spin coefficient $\gamma=\frac{A}{\sqrt2}$. 
One can  set it equal to zero by  performing a class III null rotation \cite{chandra} of the tetrad (\ref{np2})
\begin{equation}
{\bf l}\to B^{-1}{\bf l},\quad 
{\bf n}\to B {\bf n},\quad 
{\bf m}\to e^{i\Lambda} {\bf m},\quad
{\bf \bar m}\to e^{-i\Lambda} {\bf \bar m},
\label{TIPOIII}
\end{equation}
with $B=-A(z+t)$ and $\Lambda=0$. 
The final NP tetrad is 
\begin{eqnarray} 
\mathbf{l}&=&\frac{1}{\sqrt{2}}[\partial_t-\partial_z],\nonumber \\ 
\mathbf{n}&=&\frac{1}{\sqrt{2}}[\partial_t+\partial_z],\nonumber \\ 
\mathbf{m}&=&\frac{1}{\sqrt{2}}[\partial_x+i\partial_y], 
\end{eqnarray} 
and because it has constant vector components, all the spin coefficients now vanish. 
As pointed out by Teukolsky \cite{teuk2}, under a generic type III null rotation the master equation remains unchanged but there will be certain factors in front of the various quantities: $\psi_0$, $\psi_4$, $\phi_2$, etc. 
In particular, in our case all field components listed in table \ref{TAB} transform as $\Psi_s^{\rm new}=B^{-s}\Psi_s^{\rm old}$ \cite{chandra,teuk2}, and consequently the master equation (\ref{TME2}) leads to
\begin{equation}
\label{TME3} 
\left( 
\frac{\partial^2}{\partial {t^2}}-\frac{\partial^2}{\partial {x^2}}-\frac{\partial^2}{\partial {y^2}}- 
\frac{\partial^2}{\partial {z^2}} 
\right)\Psi_s^{\rm new}=0\,, 
\end{equation} 
as expected.
It is then clear from the correspondence between (\ref{TME1}) and (\ref{TME3}) that the solutions of these equations are related as well.
Finally, it is useful to express (\ref{eq:66}) in a form that would be reminiscent of black hole perturbation equations. 
Setting
\begin{equation}
P_s(Z^*)=e^{sAZ^*}a_s(Z^*) ,
\end{equation}
where $Z^*$ is the tortoise coordinate defined in equation (\ref{eq:37}),
the equation for $a_s(Z^*)$ can be obtained form (\ref{eq:66}) and is given by 
\begin{equation}
\label{tort}
\frac{\rmd^2 a_s}{\rmd Z^*{}^2}-[(k_X^2+k_Y^2) e^{-2AZ^*}+A^2\sigma^2]a_s=0.
\end{equation}
This is of the form of a one-dimensional Schr\"odinger equation and for $k_X^2+k_Y^2>0$ contains an effective potential
\begin{equation}
\mathcal{W}=-e^{-2AZ^*},
\end{equation}
which is attractive and independent of the spin of the field.

\section{The C-metric}

The vacuum C-metric, in its simplest interpretation, 
is the metric representing the exterior gravitational field of a 
uniformly accelerating Schwarzschild source. It is a member of a class of degenerate metrics 
discovered by Levi-Civita in 1918 \cite{LC}. It has been rediscovered by many authors 
\cite{newtam,robtra,ehlkun}. 
In the classification scheme of \cite{ehlkun}, it was designated as the \lq\lq C-metric". The C-metric can 
be generalized to include electric charge and angular momentum. The charged C-metric has been 
extensively studied by Kinnersley and Walker \cite{kin69,kinwal}. As described in \cite{kinwal,bon83}, the vacuum 
C-metric may represent the exterior field of one, or, by way of a certain extension, 
two uniformly accelerating spherically symmetric sources. A detailed description of 
the C-metric and a more complete list of references is available in \cite{ES}. 
The C-metric is 
of type D in the Petrov classification, and is a member of the Weyl class of solutions of 
Einstein's equations. It contains two hypersurface-orthogonal Killing vectors. 
One of these vectors 
is timelike in the static spacetime regime of interest here.
The C-metric may be expressed in the form \cite{kin69,kinwal}
\begin{equation}
\label{met_txyz}
\rmd s^2=\frac{1}{A^2(\tilde x+\tilde  y)^2}[(\tilde F \rmd t^2 - \tilde F^{-1}\rmd \tilde y{}^2) -(\tilde G^{-1} \rmd \tilde x{}^2 + \tilde G\rmd \tilde z{}^2)],
\end{equation}
where
\begin{eqnarray}
\label{eq:B2}
\fl\quad \tilde F(\tilde y):= -1+\tilde y{}^2-2mA\tilde y{}^3,\quad  \tilde G(\tilde  x):= 1-\tilde x{}^2-2mA\tilde x{}^3 . 
\end{eqnarray}

The coordinates $\{ t, \tilde x, \tilde y, \tilde z \}$ are adapted to the hypersurface-orthogonal Killing vector $\kappa=\partial_t$, 
the spacelike Killing vector $\partial_{\tilde z}$ and $\partial_{\tilde x}$, which points along the 
nondegenerate eigenvector of the hypersurface Ricci curvature. In equations (\ref{met_txyz}) and (\ref{eq:B2}), 
$m\ge 0$ and $A \ge 0$ characterize respectively the mass and acceleration of the source. 
To maintain the signature of the metric (\ref{met_txyz}), we must have $\tilde G > 0$. Moreover, 
we assume that $\tilde F > 0$. In the physical region of interest here, $mA<1/(3\sqrt{3})$ \cite{Farh,pavda,podol}.

The relation of the C-metric with the Schwarzschild metric can be elucidated by introducing the retarded coordinate $u$, the radial coordinate $r$ and the azimuthal coordinate $\phi$:
\begin{equation}
u=\frac1A [t+\int^{\tilde y} \tilde F^{-1}\rmd \tilde y], \qquad r=\frac{1}{A(\tilde  x+\tilde  y)}, \qquad \phi= \tilde z .
\end{equation}
The C-metric can then be represented as
\begin{equation}
\label{cmetEF}
\rmd s^2= \tilde H \rmd u^2 + 2 \rmd u \rmd r + 2A r^2 \rmd u \rmd \tilde x -\frac{r^2}{\tilde G} \rmd \tilde x{}^2 -r^2\tilde G \rmd \phi^2 , 
\end{equation}
where
\begin{equation}\fl\quad
\tilde H(r,\tilde x)=1-\frac{2m}{r}-A^2r^2 (1-\tilde x{}^2-2mA\tilde x{}^3)-Ar(2\tilde x+6mA\tilde x{}^2)+6mA\tilde  x .
\end{equation}
The gravitational potential $\tilde H$ determines the norm
of the hypersurface-orthogonal Killing vector $\kappa$, 
\begin{equation}
\kappa_\alpha \kappa^\alpha = r^2 \tilde F=\frac{\tilde H}{ A^2} ,
\end{equation}
hence $\kappa$ is timelike for $\tilde H>0$.

To explore some of the physical characteristics of the C-metric,
it is interesting to evaluate in this case the completely symmetric and traceless Bel-Robinson tensor
\begin{equation}\fl\quad
B_{\mu\nu\rho\sigma}=\frac12 (C_{\mu\xi\rho\zeta}C_\nu{}^{\xi}{}_\sigma{}^\zeta + C_{\mu\xi\sigma\zeta}C_\nu{}^{\xi}{}_\rho{}^\zeta )-\frac{1}{16}g_{\mu\nu}g_{\rho\sigma} C_{\alpha\beta\gamma\delta}C^{\alpha\beta\gamma\delta}.
\end{equation}
The gravitational stress-energy tensor is given by \cite{masetal1,masetal2,masetal3}
\begin{equation}
T_{(\alpha )(\beta )}=L^2 B_{\mu\nu\rho\sigma} \lambda^\mu{}_{(\alpha)}\lambda^\nu{}_{(\beta)}
\lambda^\rho{}_{(0)}\lambda^\sigma{}_{(0)}, 
\end{equation}
where $\lambda^\mu{}_{(\alpha)}$ is the orthonormal tetrad frame of the observer. 
Here $L$ is a characteristic lengthscale associated with the source; in the case under consideration, $L$ could be $L_A$, $m$ or an appropriate length formed from a combination of these.

For the C-metric in the form (\ref{met_txyz}), the orthonormal tetrad frame of the standard static observer is given by

\begin{eqnarray}
\fl\quad
\lambda{}_{(0)}=\frac{1}{r\tilde F^{1/2}}\partial_t,\quad 
\lambda{}_{(1)}= \frac{\tilde G^{1/2}}{r}\partial_{\tilde x}, \quad
\lambda{}_{(2)}=\frac{\tilde F^{1/2}}{r}\partial_{\tilde y}, \quad
\lambda{}_{(3)}= \frac{1}{r\tilde G^{1/2}}\partial_{\tilde z},
\end{eqnarray}
where $1/r=A(\tilde x + \tilde y)$.
The  gravitational stress-energy tensor measured by the static observers is 

\begin{equation}
T_ {(\alpha )(\beta )}= \epsilon \, 
\pmatrix{
      1&0&0&0\cr
	0&-1/3&0&0\cr
	0&0&2/3&0\cr
      0&0&0&2/3\cr\
}, 
\end{equation}
where $\epsilon$ is given by 
\begin{equation}
\epsilon= \frac{6m^2L^2}{r^6}.
\end{equation}
It follows from this result that there is no flow of gravitational energy according to the static observer family, in agreement with the 
static nature of the C-metric in the region under consideration in this paper.
The vacuum C-metric is Ricci-flat; therefore, it has four standard real curvature invariants given by 
\begin{eqnarray}
      I_1 &=&C_{\mu\nu\rho\sigma}C^{\mu\nu\rho\sigma}-iC_{\mu\nu\rho\sigma}{}^* C^{\mu\nu\rho\sigma} ,  \nonumber \\ 
      I_2 &=&C_{\mu\nu\rho\sigma}C^{\rho\sigma\alpha\beta}C_{\alpha\beta}{}^{\mu\nu}+i C_{\mu\nu\rho\sigma}C^{\rho\sigma\alpha\beta}{}^* C_{\alpha\beta}{}^{\mu\nu},   
\end{eqnarray}
where an asterisk denotes the dual tensor.
It turns out that in this case $I_1$ and $I_2$ are real and are given respectively by the Kretschmann scalar 
\begin{equation}
      I_1 = \frac{48 m^2}{r^6}= 8 \frac{\epsilon}{L^2} ,  
\end{equation}
and 
\begin{equation}
      I_2 = - \frac{96 m^3}{r^9}= -\frac{16}{\sqrt{6}} \frac{\epsilon^{3/2}}{L^3}=-\frac{1}{2\sqrt{3}}I_1^{3/2} .           
\end{equation}
These curvature invariants coincide with the corresponding ones in the Schwarzschild spacetime and diverge at the black hole singularity ($r\to 0$). On the other hand, 
the presence of the acceleration $A$ is reflected instead 
in the first quadratic differential invariant of the Weyl tensor
\begin{equation}
\mathcal{D}=C_{\alpha\beta\gamma\delta ;\eta}C^{\alpha\beta\gamma\delta;\eta}= -15 A^2 (\tilde F +\tilde G) I_1 .
\end{equation}
This quantity, with the metric written in the form (\ref{cmetEF}), can be expressed as
\begin{equation}
\mathcal{D}=-\frac{720 m^2}{r^8}\left[ 1-\frac{2m}{r}-2Ar\cos \theta (1-\frac{3m}{r}+3mA \cos \theta)\right],
\end{equation}
which for $A=0$ reduces to the corresponding result in the Schwarzschild spacetime.

\end{document}